\documentstyle{article}
\begin{document}

\begin{center}
{\LARGE \bf Photometry of the three eclipsing novalike variables EC~21178-5417, 
GS~Pav and V345~Pav}\footnote{Based 
on observations taken at the Observat\'orio do Pico dos Dias / LNA}

\vspace{1cm}

{\Large \bf Albert Bruch}

\vspace{0.5cm}
Laborat\'orio Nacional de Astrof\'{i}sica, Rua Estados Unidos, 154, \\
CEP 37504-364, Itajub\'a - MG, Brazil
\vspace{1cm}

(Published in: New Astronomy, Vol.\ 56, p.\ 60 -- 70 (2017))
\vspace{1cm}
\end{center}

\begin{abstract}
As part of a project to better characterize comparatively bright, yet little
studied cataclysmic variables time resolved photometry of the three eclipsing
novalike variables EC~21178-5417, GS~Pav und V345~Pav is presented. Previously
known orbital periods are significantly improved and long-term ephemeris are
derived. Variations of eclipse profiles, occurring on time scales of days to
weeks, are analyzed. Out of eclipse the light curves are characterized by low
scale flickering superposed on more gradual variations with amplitudes limited
to a few tenths of a magnitude and profiles which at least in EC~21178-5417
and GS~Pav roughly follow the same pattern in all observed cycles. 
Additionally, signs for variations on the time scale of some tens of minutes
are seen in GS Pav, most clearly in two subsequent nights when in the first
of these a signal with a period of 15.7~min was observed over several 
hours. In the second night variations with twice this period were seen. While 
no additional insight could be gained on quasi periodic oscillations (QPOs) 
and dwarf nova oscillations in EC~21178-5417, previously detected by 
Warner et al.\ (2003), and while such oscillations could not be found in 
V345~Pav, stacked power spectra of GS~Pav clearly reveal the presence of QPOs 
over time intervals of several hours with periods varying between 200 sec and 
500 sec in that system. 

\phantom{.}

{\parindent0em Keywords:
Stars: binaries: close -- 
Stars: novae, cataclysmic variables --
Stars: individual: EC~21178-5417 --
Stars: individual: GS~Pav -- 
Stars: individual: V345~Pav}
\end{abstract}

\section{Introduction}
\label{Introduction}

Cataclysmic variables (CVs) are interactive binaries where a late type, 
low mass star which is normally on or close to the main sequence transfers
matter to a white dwarf. The number of known systems of this kind has grown 
enormously in recent years mainly due to numerous detections of CVs in
large scale surveys. However, most of these newly detected systems
are rather faint in their normal brightness state. Therefore, the
characterization of their individual properties is expensive because is 
requires large telescopes.

On the other hand, it is much easier to perform detailed studies of the
brighter CVs, most of which are known for a long time. It is therefore
surprising that even among these stars an appreciable number has not
yet been adequately characterized to be certain about basic parameters. 
In some cases even the very class membership is not confirmed. 

I therefore started a small observing program aimed at a better understanding
of these so far neglected stars and to fill in evident gaps in our knowledge
about them. For this purpose I selected a number of
little studied southern CVs and suspected CVs bright enough   
to be easily observed with comparatively small
telescopes. The emphasize of this program lies on photometry with a high
time resolution aimed at the detection and characterization of short and 
medium time scale variations such as flickering and orbital variability. 

Here, I present a photometric study of three eclipsing novalike variables, 
namely EC~21178-5417, GS~Pav and V345~Pav
in order to derive improved orbital 
ephemeris and to characterize in some detail the light curves in and out of 
eclipse. In Sect.~\ref{Observations and data reductions} the observations
and data reduction techniques are described. In Sects.~\ref{EC 21178-5417}
-- \ref{V345 Pav} the results for the individual target stars are presented.
In each case, eclipse timings are measured and ephemeris are calculated, 
the eclipse profiles are studied and the out-of-eclipse variations on
various time scales are analyzed. Finally, a short summary in 
Sect.~\ref{Conclusions} concludes this paper.

\section{Observations and data reductions}
\label{Observations and data reductions}

The target stars were observed with the 0.6-m Zeiss and 0.6-m Boller \& Chivens
telescopes of the 
Observat\'orio do Pico dos Dias (OPD), operated by the Laborat\'orio Nacional
de Astrof\'{\i}sica, Brazil. During various observing missions, each lasting
a couple of nights,
time series imaging of the field around the target stars was performed
using cameras of type Andor iXon EMCCD DU-888E-C00-\#BV
equipped with back illuminated, visually optimized CCDs.
In order to resolve the expected rapid flickering variations the integration
times were kept short. Together with the small readout times of the detectors
this resulted in a time resolution of the order of 5~sec. In order to 
maximize the count rates in these short exposure intervals no filters were
used (i.e., the observations were taken in integral ``white'' light). 
Therefore, it was not possible to calibrate the stellar magnitudes.
Instead, the brightness is expressed as the magnitude difference between 
the target and a nearby comparison star. This is not a severe limitation 
in view of the purpose to the observations. However, even so the 
differential magnitudes can be roughly transformed into $V$, considering that 
the white light passband has an effective wavelength close to that of the 
$V$ band (Bruch 2017). To do so requires knowledge of the brightness of
the comparison stars. The average of their magnitudes, taken from the SPM4
(Girard et al. 2011), Nomad (Zacharias et al. 2005) and UCAC4 
(Zacharias et al. 2013)
catalogues is: $V = 12.40 \pm 0.06$ (EC~21178-5417), $V = 14.38 \pm 0.19$ 
(GS~Pav) and $V = 12.14 \pm 0.17$ (V345~Pav).  

During the various missions different CCDs had to be used. It must
be expected that their characteristics are not identical. Although the
differences should not be drastic because the CCDs were all of the
same type, I find significantly varying magnitude differences between the 
comparison stars and several check stars between missions, while during a 
given mission they remained virtually constant. This behaviour may be
attributed to slight differences of the wavelenth dependent detector
sensitivity, enhanced by the broad white light passband.
Therefore, the differential magnitudes between the target stars and
the comparison stars may be compared within a given observing run, but any
differences between missions are not reliable. This holds true for EC~21178-5417
and V345~Pav but not for GS~Pav in which case the magnitude differences between
comparison and check stars remained quite stable\footnote{All light 
curves of GS~Pav except one were obtained after conclusion of the observations
of the other two stars, using CCDs which are very similar to each other.}.

A summary of the photometric observations is 
given in Table~\ref{Journal of photometric observations}. 

\begin{table}
   \centering

\caption{Journal of photometric observations}
\label{Journal of photometric observations}

\hspace{1ex}

\begin{tabular}{llcccc}
\hline
Name      & Obs. Date & Start & End    & Time      & Number             \\
          &           & (UT)  & (UT)   & Res. (s)  & of Integr.     \\
\hline
EC 21178-5417 & 2015 Jul 14 & \phantom{2}3:04 & \phantom{2}9:02
          & \phantom{1}5\phantom{.5}  &           3\,922 \\ 
          & 2015 Aug 10/11&     23:21 & \phantom{2}7:23
          & \phantom{1}5\phantom{.5}  &           4\,961 \\ 
          & 2015 Aug 12 & \phantom{2}3:00 & \phantom{2}7:10
          & \phantom{1}6\phantom{.5}  &           3\,000 \\ 
          & 2015 Aug 13 & \phantom{2}2:40 & \phantom{2}6:48
          & \phantom{1}6\phantom{.5}  &           3\,000 \\ 
          & 2015 Aug 14 & \phantom{2}2:49 & \phantom{2}6:57
          & \phantom{1}6\phantom{.5}  &           2\,999 \\ 
          & 2015 Aug 15 & \phantom{2}2:47 & \phantom{2}6:50
          & \phantom{1}6\phantom{.5}  &           2\,125 \\ 
          & 2016 May 10 & \phantom{2}6:57 & \phantom{2}7:00
          & \phantom{1}5\phantom{.5}  & \phantom{2\,}660 \\ 
          & 2016 May 12 & \phantom{2}5:51 & \phantom{2}8:08
          & \phantom{1}5\phantom{.5}  &           1\,531 \\ [1ex]
GS Pav    & 2004 Aug 16/17* &       22:52 & \phantom{2}1:49
          & 15 -- 20                  & \phantom{2\,}474 \\ 
          & 2016 Jun 09 & \phantom{2}5:59 & \phantom{2}8:22
          & \phantom{1}5\phantom{.5}  &           2\,100 \\ 
          & 2016 Jun 28 & \phantom{2}3:05 & \phantom{2}8:54
          & \phantom{1}5\phantom{.5}  &           3\,989 \\ 
          & 2016 Jun 29 & \phantom{2}1:31 & \phantom{2}2:46
          & \phantom{1}5\phantom{.5}  & \phantom{2\,}406 \\ 
          & 2016 Aug 08/09 & 22:58        & \phantom{2}4:35
          & \phantom{1}5\phantom{.5}  &           3\,227 \\ 
          & 2016 Aug 11 & \phantom{2}1:15 & \phantom{2}5:26
          & \phantom{1}5\phantom{.5}  &           3\,000 \\ 
          & 2016 Aug 12 & \phantom{2}0:46 & \phantom{2}5:36
          & \phantom{1}5\phantom{.5}  &           3\,495 \\ 
          & 2016 Sep 06 & \phantom{2}1:48 & \phantom{2}3:40
          & \phantom{1}5\phantom{.5}  &           1\,346 \\ 
          & 2016 Sep 06/07 &        22:21 & \phantom{2}3:18
          & \phantom{1}5\phantom{.5}  &           1\,863 \\ [1ex] 
V345 Pav  & 2015 May 22 & \phantom{2}5:36 & \phantom{2}7:13
          & \phantom{1}6\phantom{.5}  & \phantom{1\,}993 \\
          & 2015 Jun 09 & \phantom{2}2:36 & \phantom{2}7:32
          & \phantom{1}6\phantom{.5}  &           2\,998 \\
          & 2015 Jun 10 & \phantom{2}1:58 & \phantom{2}5:13
          & \phantom{1}6\phantom{.5}  &           2\,000 \\
          & 2015 Jun 11 & \phantom{2}1:50 & \phantom{2}5:30
          & \phantom{1}6\phantom{.5}  &           1\,997 \\
          & 2015 Jun 12 & \phantom{2}1:49 & \phantom{2}5:04
          & \phantom{1}6\phantom{.5}  &           2\,000 \\ 
          & 2015 Aug 10 &       21:36 &            22:59
          & \phantom{1}5\phantom{.5}  &           1\,000 \\ 
          & 2015 Aug 11/12&     21:24 & \phantom{2}2:39
          & \phantom{1}5\phantom{.5}  &           3\,495 \\ 
          & 2015 Aug 12/13&     21:28 & \phantom{2}2:29
          & \phantom{1}5\phantom{.5}  &           3\,596 \\ 
          & 2015 Aug 13/14&     21:27 & \phantom{2}2:30
          & \phantom{1}5\phantom{.5}  &           3\,494 \\ 
          & 2015 Aug 14/15&     21:57 & \phantom{2}2:37
          & \phantom{1}5\phantom{.5}  &           3\,303 \\
          & 2016 Jun 09&\phantom{2}4:09 & \phantom{2}5:50
          & \phantom{1}5\phantom{.5}  &           1\,158 \\
          & 2016 Jun 10&\phantom{2}4:03 & \phantom{2}5:25
          & \phantom{1}5\phantom{.5}  &           1\,000 \\ [1ex]
\hline
\multicolumn{6}{l}{*archival data, filter $B$ (unpublished)}
\end{tabular}
\end{table}
%

\begin{figure}
   \parbox[]{0.1cm}{\epsfxsize=14cm\epsfbox{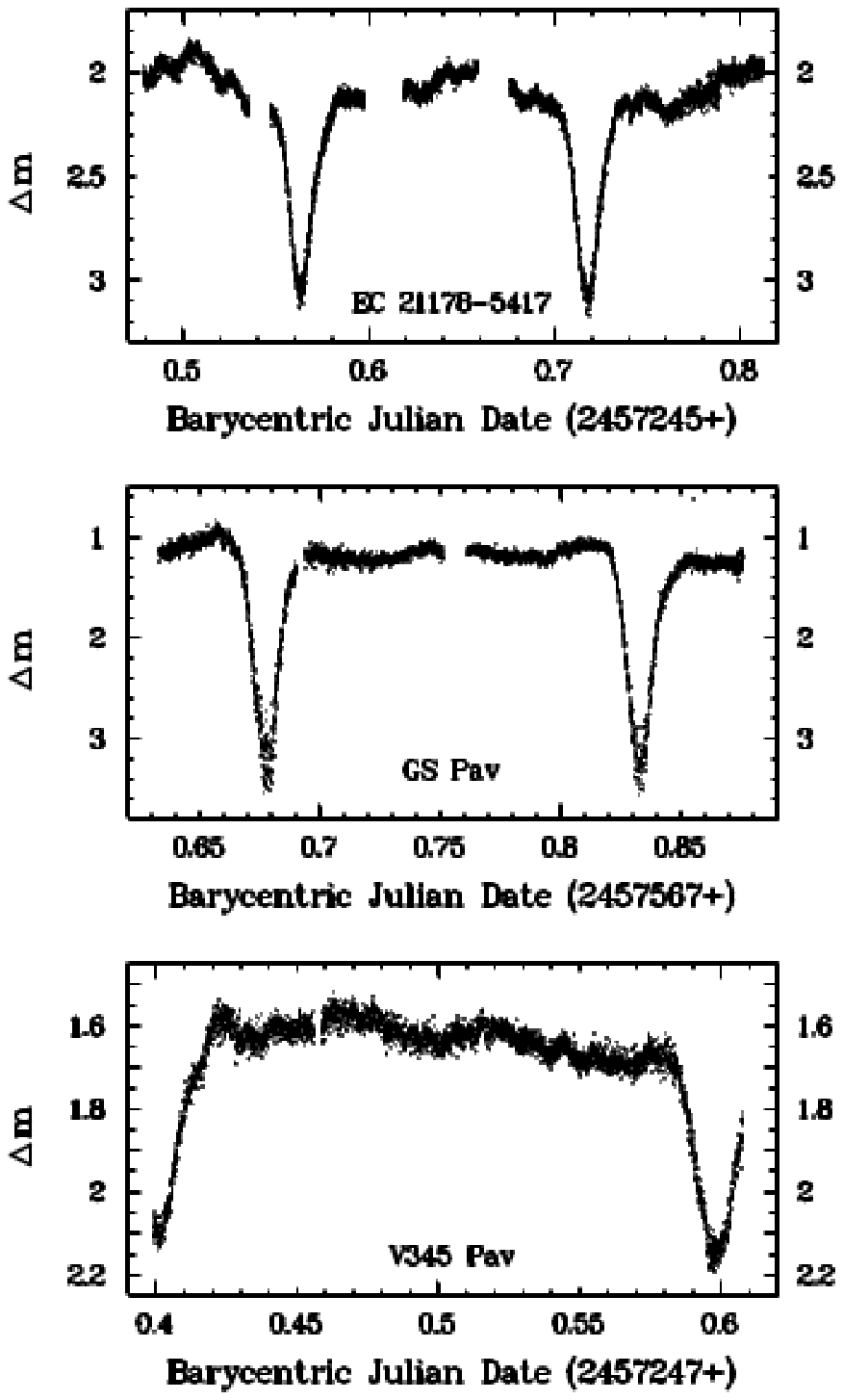}}
      \caption[]{Light curves of EC~21178-5417 (top), GS~Pav (centre) and 
                 V345~Pav (bottom)
                 as examples of typical orbital light curves of these 
                 systems.}
\label{example-lc}
\end{figure}

Basic data reductions (biasing, flat-fielding) was performed using IRAF. 
For the construction of light curves aperture photometry routines 
implemented in the MIRA software system (Bruch 1993) were employed. The
same system was used for all further data reductions and calculations. 
All timing information was transformed into barycentric Julian Dates
on the barycentric dynamical time scale (BJD-TDB), 
following Eastman et al.\ (2010). Examples of the light curves are shown
in Fig.~\ref{example-lc}.

\input epsf

\section{EC~21178-5417}
\label{EC 21178-5417}

Not much is known about EC~21178-5417. Identified as a blue object in the
Edinburgh Cape survey (Stobie et al. 1997), it was first mentioned as a 
cataclysmic
variable by Warner et al.\ (2003) who found it to be an eclipsing novalike
system and quote an orbital period of 3.708~h. Since that paper 
was exclusively concerned with dwarf nova oscillations (DNOs) and 
quasi-periodic oscillations (QPOs) in a larger sample of CVs, 
Warner et al.\ (2003) did not investigate further properties of 
EC~21178-5417 apart from that particular aspect. They postponed details 
about the light curve and the spectrum to another paper which, however, 
never appeared.

\subsection{Eclipse timings and ephemeris}
\label{EC Eclipse timings and ephemeris}

The light curves contain a total of 10 eclipses. The minimum times were 
determined by the time of minimum $t_{\rm m}$ of a Gaussian and a polynomial 
fit to the data within a range of $\pm$10~min around the minimum. Changing 
the degree of the polynomial has only a minute influence of the results. In
particular, there are no systematic variations with the polynomial degree.
Choosing all values between 2 and 9 and adding the minimum of the fitted 
Gaussian resulted in an average standard deviation of 10.8~sec for $t_{\rm m}$ 
of the 10 eclipses. Thus, this value may be regarded as the typical error of 
the eclipse time determination. 

I adopted the average values of $t_{\rm m}$ for the different polynomial degrees
and the Gaussian as the time of minimum of eclipses. They are listed in 
Tab.~\ref{Eclipse timings} where eclipse no.\ \#0 is arbitrarily assigned 
to the first eclipse observed on 2015, August 11.   

\begin{table}
   \centering

\caption{Eclipse epochs}
\label{Eclipse timings}

\hspace{1ex}

\begin{tabular}{rlr|rlr|rlr}
\hline
\multicolumn{3}{c|}{EC~21178-5417} & 
\multicolumn{3}{c|}{GS~Pav}        & 
\multicolumn{3}{c}{V345 Pav}       \\
Ecl. No. & BJD-TDB & \multicolumn{1}{c|}{$O-C$}  &
Ecl. No. & BJD-TDB & \multicolumn{1}{c|}{$O-C$}  &
Ecl. No. & BJD-TDB & \multicolumn{1}{c}{$O-C$}   \\
         & (2400000+) & \multicolumn{1}{c|}{(sec)} &
         & (2400000+) & \multicolumn{1}{c|}{(sec)} &
         & (2400000+) & \multicolumn{1}{c}{(sec)}  \\
\hline
  -180 & 57217.74883 &  31.0 & 
-53662 & 49235.73769 & -35.7 &
-47429 & 47769.28105 &  24.8 \\
     0 & 57245.56294 & -42.9 & 
-53650 & 49237.60208 &  64.6 &
-47428 & 47769.47865 & -18.1 \\
     1 & 57245.71788 &  -7.3 & 
-53644 & 49238.53368 &  62.8 &
-47419 & 47771.26145 & -23.9 \\
     7 & 57246.64511 &  -1.7 & 
-53643 & 49238.68828 &   4.9 &
-47418 & 47771.45985 &   2.3 \\
     8 & 57246.79950 & -13.8 & 
-50035 & 49798.90168 & -36.7 &
-47414 & 47772.25245 &  20.8 \\
    14 & 57247.72708 &  22.4 & 
-50029 & 49799.83468 &  82.6 &
-47287 & 47795.42915 & -29.1 \\
    20 & 57248.65353 & -39.9 & 
-49502 & 49881.66098 &   0.4 &
-47287 & 47797.41065 &  17.3 \\
    27 & 57249.73531 & -32.5 & 
-49501 & 49881.81538 & -74.7 &
     0 & 57164.79457 &  20.6 \\
  1768 & 57518.76817 &  -3.1 & 
-49321 & 49909.76478 &  -4.4 &
    90 & 57182.62304 &   3.1 \\
  1781 & 57520.77724 &  15.3 & 
-49320 & 49909.91978 & -27.8 &
    95 & 57183.61366 &  14.8 \\
       &             &       & 
-49231 & 49923.73828 & -73.0 &
   100 & 57184.60394 &  -2.5 \\
       &             &       & 
-49089 & 49945.78638 & -92.8 &
   105 & 57185.59431 & -12.7 \\
       &             &       & 
-48896 & 49975.75488 &  28.5 &
   407 & 57245.41978 &  18.8 \\
       &             &       & 
-48871 & 49979.63688 &  50.3 &
   412 & 57246.41023 &  16.1 \\
       &             &       & 
-27908 & 53234.56104 & 116.0 &
   413 & 57246.60815 &   0.9 \\
       &             &       & 
  -116 & 57549.82129 & -10.4 &
   417 & 57247.40095 &  36.7 \\
       &             &       & 
    -1 & 57567.67787 &  36.1 &
   418 & 57247.59809 & -46.4 \\
       &             &       & 
     0 & 57567.83294 &  18.6 &
   423 & 57248.58944 &  28.8 \\ 
       &             &       & 
     5 & 57568.60883 & -21.2 &
   428 & 57249.57976 &  14.9 \\
       &             &       & 
   269 & 57609.60029 &  -3.5 &
  1938 & 57548.70474 & -33.5 \\
       &             &       & 
   282 & 57611.61918 &  26.0 &
  1943 & 57549.69514 & -41.2 \\
       &             &       & 
   288 & 57612.55037 &  -8.4 &
       &             &       \\
       &             &       & 
   289 & 57612.70612 &  33.2 &
       &             &       \\
\hline
\end{tabular}
\end{table}

A linear least squares fit to the eclipse epochs in 
Tab.~\ref{Eclipse timings}, weighted by the inverse 
of the error of the individual timings, yields the ephemeris
\begin{eqnarray}
{\rm BJD_{min}} & = & 2457245.56348\ (10) \nonumber \\ [-0.5ex]
               &    & + 0.15452758\ (14) \times E \nonumber 
\end{eqnarray}
where $E$ is the cycle number. The errors are given in units of the last 
decimal digits. They are formal errors of the least squares fit which may
sub estimate the real errors. The rms of the residuals between
the observed eclipse times and those calculated from the above ephemeris
(i.e., the $O-C$ values) is 25.7~sec. The $O-C$ values for the
individual eclipses are included in Table~\ref{Eclipse timings}.

\subsection{Eclipse profile}
\label{EC Eclipse profile}

\begin{figure}
   \parbox[]{0.1cm}{\epsfxsize=14cm\epsfbox{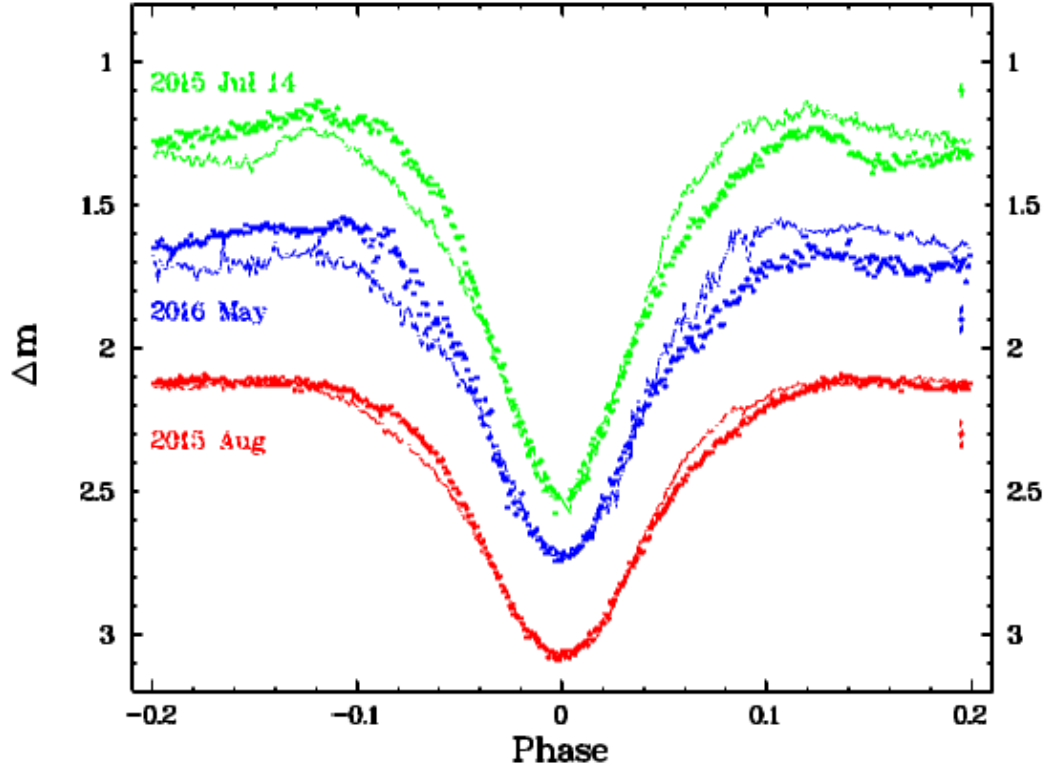}}
      \caption[]{Eclipse profile of EC~21178-5417 on 2015, July 14
                 (green) and average profile of seven (red) and two (blue) 
                 eclipses observed in 2015, August and 2016, May, respectively.
                 All light curves were binned in phase intervals of
                 width 0.001. Error bars are indicated at the right margin
                 above or below the curves. The thin continuous curves are 
                 the eclipse profiles mirrored at phase 0 in order to emphasize
                 asymmetries in their upper part. 
                 (For visualization of the colours used in this figure, the
                 reader is referred to the web version of this article.)}
\label{ec-eclprof}
\end{figure}

Fig.~\ref{ec-eclprof} shows the eclipse profiles of EC~21178-5417 as a 
function of orbital phase. The upper (green) graph refers to the 2015,
July 14, while the others show the average of all 2015, August (red) and
2016, May (blue) light curves. The latter has been shifted downward by
$0^{\raisebox{.3ex}{\scriptsize m}}_{\raisebox{.6ex}{\hspace{.17em}.}}2$ for clarity.
In all cases the data were binned in intervals of width 0.001 in phase. 
The average of the standard deviations of the individual magnitudes in
each interval is shown as a typical error bar at the right margin just below
or above the light curves. It is larger in 2015, August in spite of the less 
noisy average profile as compared to that of 2015, July 14 because of 
slight systematic differences between the seven contributing eclipses. The same
is true for 2016, May, where additionally parts of the light curves (but
fortunately not the eclipse bottom) were affected by passing clouds, increasing
the noise.

While the lower part of the eclipse profiles 
is symmetrical to a high degree, slight deviations from
symmetry occur during the initial and final parts, respectively, of ingress
and egress. This is more obvious when regarding the thin continuous
curves in Fig.~\ref{ec-eclprof}
which are the eclipse profiles after being mirrored at phase 0. The first part
of the ingress is slightly steeper than the last part of the egress. This
effect is seen (often much stronger than in the present case) in many other
eclipsing cataclysmic variables and is generally attributed to different 
aspects of the mass transfer stream impact region on the accretion disk at 
the start and end of eclipse.

In order to estimate the eclipse depth the magnitude difference 
$\Delta m_{\rm ecl}$ between the eclipse bottom (defined as the minimum of a 
high degree polynomial fit to the phase interval --0.04 \ldots 0.04) and the 
mean magnitude $\Delta m$ in the phase intervals --0.20 \ldots --0.18 and 
0.18 \ldots 0.20, respectively, was calculated.
The results are listed in Table~\ref{eclipse depth}.
It is seen that the depth
remains fairly constant at a given epoch but differs significantly on longer
time scales. In particular, the eclipses were shallower during 2015, August
than in 2015, July. 

\begin{table}
   \centering

\caption{Eclipse depth}
\label{eclipse depth}

\hspace{1ex}

\begin{tabular}{rl|rl|rl}
\hline
\multicolumn{2}{c|}{EC~21178-5417} & 
\multicolumn{2}{c|}{GS Pav} &
\multicolumn{2}{c}{V345 Pav} \\
Ecl.& depth  & Ecl. & depth & Ecl. & depth \\
No. & (mag)  & No.  & (mag) & No.  & (mag) \\
\hline
-180 & 1.22 & -27908 & \phantom{$<$}3.85*           &  0   & 0.76 \\  
   0 & 0.92 &   -129 & \phantom{$>$}2.23\phantom{*} &  90  & 0.75 \\
   1 & 0.97 &     -1 & \phantom{$>$}2.19\phantom{*} &  95  & 0.74 \\
   7 & 0.93 &      0 & \phantom{$>$}2.08\phantom{*} &  100 & 0.75 \\
   8 & 0.96 &      5 & \phantom{$>$}2.39:           &  105 & 0.74 \\
  14 & 0.99 &    269 & \phantom{$>$}2.00\phantom{*} &  407 & 0.42 \\
  20 & 0.93 &    282 & \phantom{$>$}2.28\phantom{*} &  412 & 0.44 \\
  27 & 0.93 &    288 & \phantom{$>$}2.28\phantom{*} &  413 & 0.50 \\
1768 & 1.10 &    289 & \phantom{$>$}2.09\phantom{*} &  417 & 0.43 \\
1781 & 1.01 &    455 &           $>$1.79\phantom{*} &  418 & 0.51 \\
     &      &        &                              &  423 & 0.47 \\
     &      &        &                              &  428 & 0.51 \\
     &      &        &                              & 1938 & 0.72 \\
     &      &        &                              & 1943 & 0.79 \\
\hline        
\multicolumn{6}{l}{*$B$ band}
\end{tabular}  
\end{table}


\subsection{Out-of-eclipses variations}
\label{EC Out-of-eclipse variations}

As observed in all CVs, the out-of-eclipse variations exhibit flickering
which in the present case is limited to amplitudes up to
about $0^{\raisebox{.3ex}{\scriptsize m}}_{\raisebox{.6ex}{\hspace{.17em}.}}1$. This is 
on the low side of the amplitude distribution in CVs, but not unusual for 
systems with accretion disks in a bright state such as novalike variables 
and old novae (Beckemper 1995). Most of the
higher frequency flickering variations are drowned in the data noise. 
Therefore, I refrain from a more thorough quantitative analysis.

\begin{figure}
   \parbox[]{0.1cm}{\epsfxsize=14cm\epsfbox{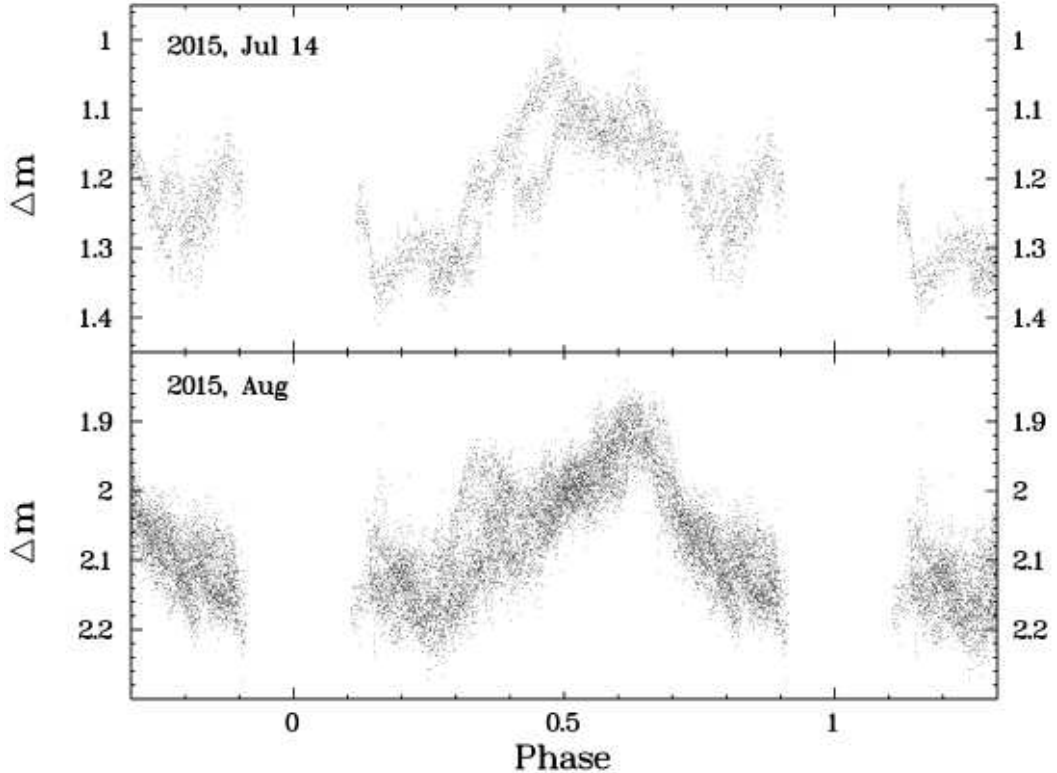}}
      \caption[]{Phase folded out-of-eclipse light curves of 
                 EC~21178-0517 of 2015, July 14 (top) and August 11 -- 15 
                 (bottom).}
\label{ec-out-of-eclipse}
\end{figure}

Superposed on the flickering are variations on longer time scales. 
In order to study their dependency on orbital phase, the eclipses were 
first removed. Thereafter, the combined 2015, August and the single 2015, July 
light curves
(the latter encompassing about 1.6 cycles) where folded on the orbital
period. The results are shown in Fig.~\ref{ec-out-of-eclipse}\footnote{The
2016, May data are not included in this comparison because the light curves
are restricted to only a limited phase range around eclipse.}. 
The difference of the average magnitude of EC~21178-5417 outside eclipse
during the two epochs is significantly larger
than can be explained as being due to the use of different detectors during
the respective observing missions 
(see Sect.\ \ref{Observations and data reductions}). Therefore, it is safe to
say that the star was really brighter in July than in August, although I
cannot quantify by how much.

During August (lower frame),
the light curve is characterized by a gradual and linear increase
of brightness between phases $\phi = 0.25$ and $\phi = 0.65$ and then a
more rapid decline until $\phi = 0.8$. The disturbance
seen at $0.3 \le \phi \le 0.4$ occurs only in 2 of six cycles with data
covering this phase range. The eclipses may hide a secondary maximum at
$-0.2 \le \phi \le 0.25$. This is better seen in the light curve of July 14
(upper frame) which also shows the primary maximum, now shifted to
earlier phases. The approximate constancy of the profile of these variations
suggest the presence of structures in the accretion disk which remain stable
at least on the time scale of weeks. 

I investigated the possible presence of consistent brightness variations 
with periods others than the orbital period $P_{\rm orb}$. For this purpose 
the combined 2015, August light curves (masking the eclipses) were subjected to 
various period
search algorithms [Lomb-Scargle periodogram, Lomb (1976),
Scargle (1982); power spectra following Deeming (1975);
phase dispersion minimization (PDM), Stellingwerf (1978); analysis of 
variance (AoV), Schwarzenberg-Czerny (1989)]. To no avail: Only 
$P_{\rm orb}$ and its harmonics as well as their aliases caused by the window
function were recovered. Pre-whitening the data, subtracting the orbital 
variations before the period analysis did not change the picture. Therefore,
I am confident that EC~21178-5417 does not exhibit periodic variations
different from $P_{\rm orb}$ which remain coherent over the several nights 
spanned by an observing mission.

\subsection{DNOs and QPOs}
\label{EC DNOs and QPOs}

Warner et al.\ (2003) identified EC~21178-5417 as a rich source of short period
oscillations. They present an extensive list of DNOs, longer period dwarf 
nova oscillations (lpDNOs) and QPOs which appear for some time in the light 
curves observed by them and 
are absent at other times. The DNOs have periods between 22~sec and 
26~sec (sometimes their first overtone is observed; they may also split up
into more than one closely spaced components). The periods of the 
lpDNO (which more often than not appear simultaneously with the DNOs) 
range between $\sim$80~sec and $\sim$100~sec while the QPOs 
span a wider range between $\sim$200~sec and $\sim$500~sec. In his
review on rapid oscillations in cataclysmic variables Warner (2004)
discusses the relationship between DNOs, lpDNOs and QPOs in EC~21178-5417 and
in other systems. 

Comparing Fig.~14 of Warner et al.\ (2003) with the present data it is
obvious that the S/N ratio of their light curves is higher. Therefore, it 
will be more difficult here to detect short period oscillations 
which may be present. Nevertheless, an effort to do so and eventually
put constraints on their amplitudes is worthwhile.

When it comes to QPOs, their low coherence and often short duration (limited
frequently to only a few cycles) in the presence of the ubiquitous flickering in
cataclysmic variables severely hampers their detection using tools such as 
Fourier transforms. As Bruch (2014) pointed out, it is not obvious
where to draw the borderline between QPOs and flickering. In view of these
difficulties, Warner (2004) wisely (citing) {\em ``accepts only the 
QPOs in the light curves that are obvious to the eye''}. Adopting this 
criterion I cannot confidently identify any QPOs in the present light curves.

\begin{figure}
   \parbox[]{0.1cm}{\epsfxsize=14cm\epsfbox{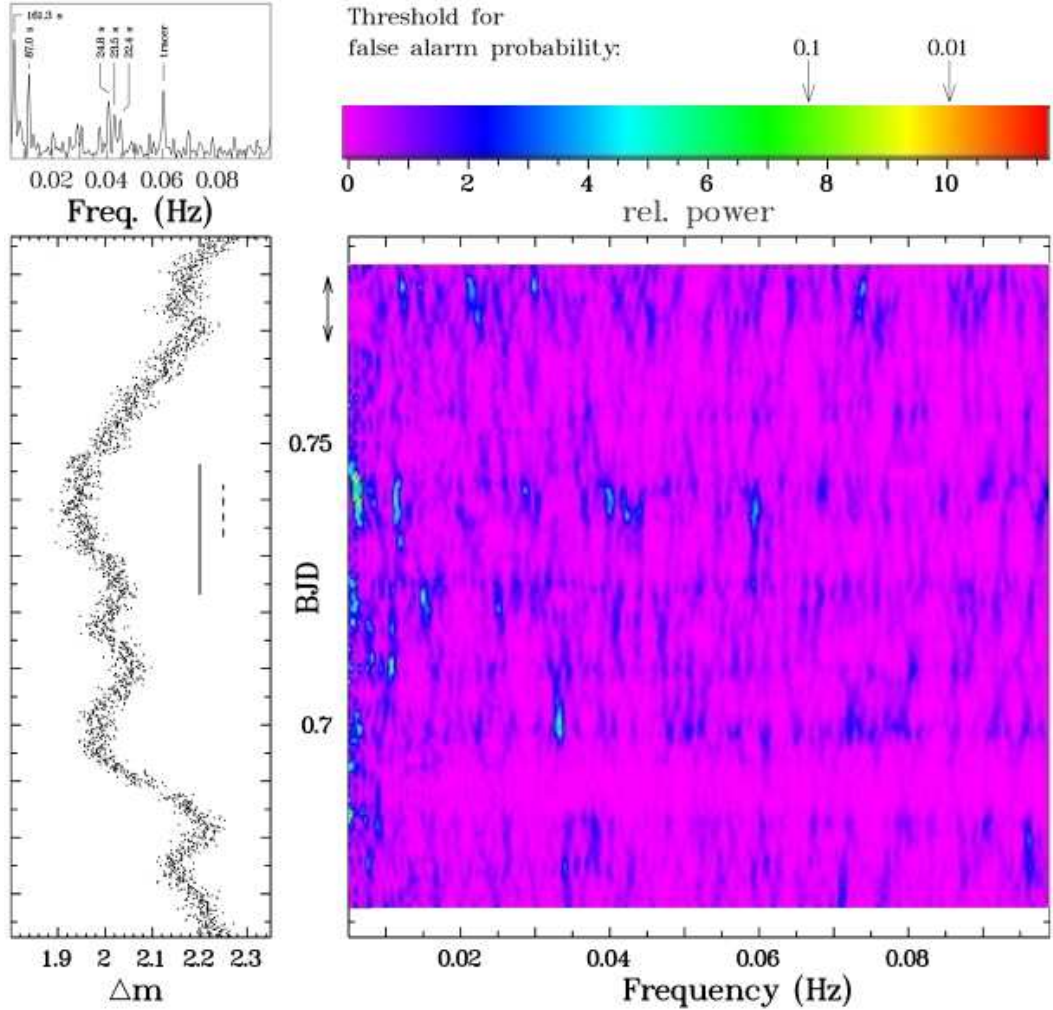}}
      \caption[]{Out-of-eclipse light curve of EC~21178-5417 of 2015, Aug. 12
                (lower left frame) and stacked power spectra of the same
                data (right). Spectral features within a range of 1000~sec 
                are not independent from each other. The length of this range
                is indicated by a double arrow at the upper left margin of the
                stacked power spectra. The levels of false alarm probabilities 
                0.1 and 0.01 are indicated on the colour bar (upper right). The 
                solid bar underneath the light curve indicates a section where 
                a tracer signal of 0.06~Hz was added. The dashed line indicates
                a range in time where DNOs and lpDNOs may be present. An
                alternative representation of the power spectrum of this range
                is shown in the upper left frame. 
                (For visualization of the colours used in this figure, the
                reader is referred to the web version of this article.)}
\label{ec-qpo}
\end{figure}

DNOs (and lpDNOs) are more coherent over the time scale of hours
[Warner (2004) cites $10^3 < | dP/dt |^{-1} < 10^7$] and are thus
more easily detected in power spectra.  However, since in many cases they are 
only present during a part of an observed light curve, they may not reveal
themselves in power spectra of the entire curve. Therefore, I calculated
stacked power spectra of all light curves, following Bruch (2014):
Lomb-Scargle periodograms for sections of a data train, 
1000~sec long, were constructed, 
allowing for an overlap of 900~sec between subsequent sections. The
individual power spectra\footnote{I use the terms ``power spectrum'' and
``periodogram'' synonymously.} were then stacked on top of each other, resulting
in a two dimensional representation (frequency vs. time). As an example,
Fig.~\ref{ec-qpo} shows the light curve of 2015, August 12 (lower left) and the
corresponding stacked power spectrum (lower right). Power is coded as shown
by the colour bar (upper right) where also false alarm probabilities 
$P_{\rm fa} = 0.1$ and $P_{\rm fa} = 0.01$ 
for peaks in the power spectra are indicated. They were calculated
using Eq. 18 of Scargle (1982), the number of independent frequencies 
having been determined as detailed by Bruch (2016). Spectra
within a time scale of 1000~sec are not independent (therefore, the length
of the sections from which the power spectra are calculated, will henceforth
be termed ``independence limit''). Thus, only structures with a 
vertical extension of $>$1000~sec (the length of the 
double arrow at the upper left of the stacked power spectra) indicate 
persistent signals. 

In order to assess the sensitivity of the power spectra to oscillations, a
sine wave with an amplitude of 
$0^{\raisebox{.3ex}{\scriptsize m}}_{\raisebox{.6ex}{\hspace{.17em}.}}005$
and a frequency of 0.06~Hz
was added as a tracer signal to the part of the light curve marked by the 
solid vertical line in the lower left frame of Fig.~\ref{ec-qpo}\footnote{The
reproduced light curve does not contain the tracer signal.}. It can be
identified in the stacked power spectra, but is not a particularly outstanding
feature. Thus, in order to confidently identify any real oscillations in the
data they should have amplitudes larger than the tracer signal. This is more
than the oscillation amplitudes observed by Warner et al.\ (2003).

Therefore, it is not surprising that 
I cannot detect strong evidence for the presence of significant oscillations.
The DNOs observed by Warner et al.\ (2003) have a frequency predominantly
just above 0.04~Hz. While during some time intervals signals are seen at 
these frequencies, by no means they stand out against many other signals
at other time intervals and frequencies. Drawing on the pre-information about
the expected frequencies and the often simultaneous presence of DNOs and lpDNOs,
the strongest evidence for the presence of oscillations occurs in the small
time interval marked by the broken vertical line in the lower left frame of
Fig.~\ref{ec-qpo}. An alternative representation of the power spectrum of this
region is shown in the upper left frame of the figure. Three peaks 
corresponding to periods of 22.4~sec, 23.5~sec and  
24.8~sec may represent DNOs split up into multiple components, while
the peak corresponding to 87.0~sec may be a lpDNO. Additionally, a 
low frequency signal (161.3~sec) is the only feature in the stacked
power spectrum which attains a false alarm probability below 0.01 
($P_{\rm fa} = 0.002$). Thus, while oscillations such as those observed by
Warner et al.\ (2003) may be present in our data, they cannot be considered
independent detections.

The stacked power spectra of the other light curves are similar to that of
2015, August 12. In only one case, during a short period of time on 2015, 
August 14, possibly real signals (however, only with marginal significance) 
were detected simultaneously at 10.6~sec and at 
161.3~sec. The former may be the overtone of a DNO at 
21.2~sec, while the latter has a period identical to the low
frequency oscillation observed on 2015, Aug. 12. 

\section{GS Pav}
\label{GS Pav}

GS~Pav was discovered as a variable star by Hoffmeister (1963).
Zwitter \& Munari (1995)
published a spectrum. Only the Balmer emission lines can 
reliably be identified upon a blue continuum. Groot et al.\ (1998) presented
the only more detailed study of GS~Pav. Their photometric measurements
reveiled the star to be deeply eclipsing with a period of 3.72 h.
They estimated masses of the components, restricted the
orbital inclination to
$74^{\raisebox{.3ex}{\scriptsize o}} < i < 83^{\raisebox{.3ex}{\scriptsize o}}$ 
and argued that 
the disk radius is variable which leads to a correlation between the eclipse
depth and the out-of-eclipse magnitude. They also classified the system to be
a novalike variable of the RW~Tri subclass.

GS~Pav was initially included in the present study because the observation of
additional eclipses and the time difference to the observations of 
Groot et al.\ (1998) should permit to improve the precision of the orbital 
period by an order of magnitude. The observations listed in 
Tab.~\ref{Journal of photometric observations}, performed within an interval
of about two months, were supplemented by archival data retrieved from the
LNA data bank, consisting of a light curve observed at the 1.6 m Perkin
Elmer telescope of OPD on 2004, August 16. These data were obtained with a
$B$ filter and span a time interval of about 3~h at a time resolution of
15~sec (first hour) and 20~sec (last two hours). 

\begin{figure}
   \parbox[]{0.1cm}{\epsfxsize=14cm\epsfbox{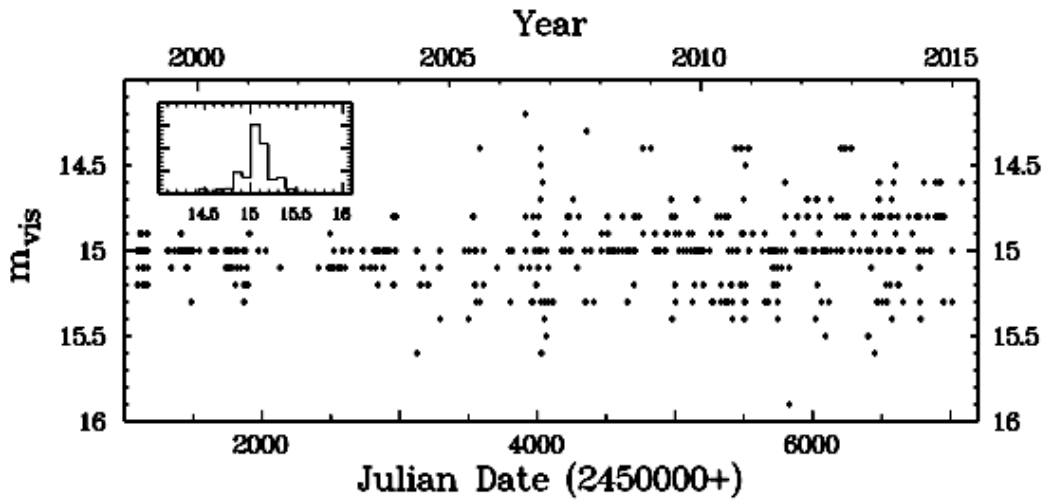}}
      \caption[]{Long term visual light curve of GS~Pav based on AAVSO
                 data. The insert shows a histogram of the magnitude 
                 values.}
\label{gspav-aavso}
\end{figure}

Ritter \& Kolb (2003) classified GS~Pav as being of the VY~Scl subtype, i.e.,
a novalike variable which occasionally drops into a low state. Although Groot
et al.\ (1998) found the out-of-eclipse magnitude of the star to vary, they
argued that GS~Pav should not be considered a VY~Scl star. This is 
corroborated by the AAVSO long term light curve shown in Fig.~\ref{gspav-aavso}
which shows that the star has a normal visual magnitude of 
$\sim$$15^{\raisebox{.3ex}{\scriptsize m}}$
with a limited scatter to higher and lower values. The coverage is dense
enough such that substantial low states cannot hide within the gaps. Therefore,
there appears to be no reason to suspect a VZ~Scl nature for GS~Pav.

\subsection{Eclipse timings and ephemeris}
\label{GS Eclipse timings and ephemeris}
 
The light curves contain a total of 9 eclipses\footnote{Here, I disregard two
eclipses observed on 2016, September 06/07. 
Observations of the first of them had
to be interrupted because of incoming clouds just about at the time of minimum.
Of the other one only the eclipse bottom -- quite noisy because of the 
faintness of the star at that phase -- could be observed. Therefore, 
in both cases it was impossible to determine the minimum time reliably.}. 
In order to determine the
minimum epochs a range in time of approximately 14~min before and 11~min
after eclipse minimum was regarded. This range encompasses the entire ingress,
but avoids the shallower part at the end of egress. Restricting this exercise
to a smaller range around the bottom of the eclipse is not warranted because
at least in some light curves these phases are quite noisy due to the faintness
of the system during mid-eclipse. The minimum times were then determined in the
same way as for EC~21178-5417. Their typical error was found to be 
19.0~sec. The average timings are listed in
Tab.~\ref{Eclipse timings} together with the eclipse times measured by 
Groot et al.\ (1998). Since the latter are given in HJD expressed on the
UTC time scale (P. Groot, private communication) they have been transformed
into BJD-TDB, using the online tool of Eastman et al.\ (2010).
Eclipse no.\ \#0 is arbitrarily assigned to the second eclipse 
observed on 2016, June 28, noting that the ephemeris of Groot et al.\ (1998) 
are sufficiently accurate to ensure cycle count continuity to the present 
epoch (large negative eclipse numbers). Eclipse number -27908 refers
to the archival light curve of 2004, August 16/17. 
Assigning weights in proportion to the inverse of the individual timing
errors to the 2016 data points and the average of these weights to the data
points of Groot et al.\ (1998), a linear least squares fit
to the eclipse epochs yields the ephemeris 
\begin{eqnarray}
{\rm BJD_{min}} & = & 2457567.83273\ (13) \nonumber \\ [-0.5ex]
               &   & + 0.1552699235\ (36) \times E \nonumber 
\end{eqnarray}
{\parindent0em The rms of the $O-C$ values is 50.6~sec in this case and
the respective values for the individual eclipses are listed in 
Table~\ref{Eclipse timings}.}

The distribution of eclipse timings consists of two groups of data points
separated by about 22 years, plus a single epoch about halfway between them.
If the period of GS~Pav had a significant derivative, the assumption of linear
ephemeries should result in a large $O-C$ value for that point. At 
116 sec it is slightly more than two times the overall rms. Thus,
any period derivative of GS~Pav must be so small that is has at most a 
marginal effect over the total time base of the present observations. 

\subsection{Eclipse profile}
\label{GS Pav Eclipse profile}

\begin{figure}
   \parbox[]{0.1cm}{\epsfxsize=14cm\epsfbox{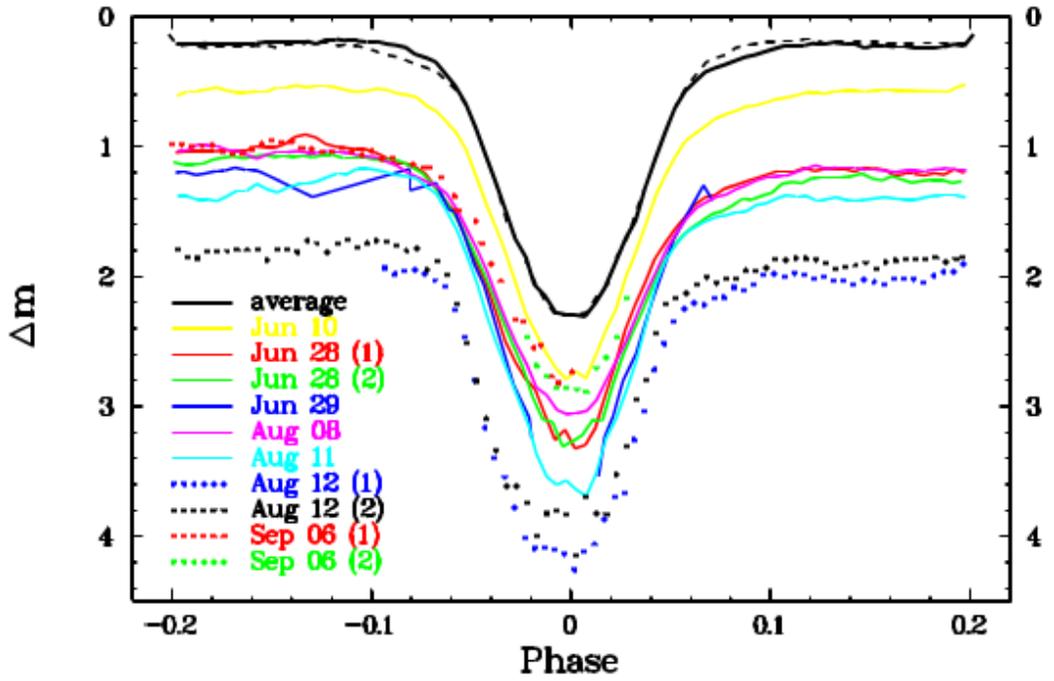}}
      \caption[]{Eclipse profiles of GS~Pav after binning the original data
                 points in intervals of width 0.005 in phase. The solid black
                 curve represents the average profile after applying a
                 magnitude correction to the individual light curves (defined
                 by the average magnitude in the phase intervals --0.20 \ldots
                 --0.18 and 0.18 \ldots 0.20) in order to take into account
                 night-to-night variations of the mean system brightness.
                 The broken back curve is a mirrored version of the 
                 average profile in order to visualize any asymmetries.
                 (For visualization of the colours used in this figure, the
                 reader is referred to the web version of this article.)}
\label{gspav-eclprof}
\end{figure}

The individual eclipses of GS~Pav, binned in phase intervals of width 0.005,
are plotted in Fig.~\ref{gspav-eclprof}. With the exception of the uppermost
curve, which represents the average profile after applying a magnitude 
correction to the individual light curves (defined by the average magnitude 
in the phase intervals --0.20 \ldots --0.18 and 0.18 \ldots 0.20) in order to 
take into account night-to-night variations of the mean system brightness, 
all data have been left on their original differential magnitude scale
(i.e., without applying vertical shifts). It is thus seen that the magnitude
level just before eclipse varies within a range of 
$1^{\raisebox{.3ex}{\scriptsize m}}_{\raisebox{.6ex}{\hspace{.17em}.}}2$. Changes of
several tenths of a magnitude can occur from night to night (e.g., between
2016, Aug. 11 and 12). The average eclipse profile is symmetrical around
phase 0 to a high degree. This is emphasized by the broken back curve
which is a mirrored version of the original profile. It only deviates from
the original at the extremes of the profile due to a delayed late egress
of the accretion disks. As in the case of EC~21178-5417 this suggest a
phase dependent contributions of the stream impact region to the total light. 

The eclipses of GS~Pav are much deeper than in the other objects of this
study. The depth $\Delta m_{\rm ecl}$ was measured in the same way as in 
EC~21178-5417. It is listed in Table~\ref{eclipse depth}.

For completeness I also measured the eclipse depth in the archival $B$ band
light curve of 2004, Aug. 16/17 and include the result in 
Table~\ref{eclipse depth}. Not surprisingly for a blue object such as a CV 
it is significantly deeper than in the white light data.

The latter correspond roughly to the $V$ band \cite{Bruch17}.
Thus, $\Delta m \approx \Delta V$ and
$\Delta m_{\rm ecl} \approx \Delta V_{\rm ecl}$ .
Since the data of Groot et al.\ (1998) also refer to the $V$ band, a 
comparison with their eclipse depth measurements can be made. They find
a complex correlation between $\Delta V_{\rm ecl}$ and $V$ (their Fig.~2). 
As was pointed out in Sect.~\ref{Observations and data reductions},
in contrast to the other two stars of this study the zero point of the 
differential magnitude scale for GS~Pav did no change between missions. It is 
therefore possible to look for a similar correlation in the present data. It 
does not exist. A formal linear least squares fit yields a correlation 
coefficient of only 0.15 between these quantities and the gradient 
$\Delta V_{\rm ecl}/\Delta V$ has a formal error which is more than twice its 
nominal value. 

The present observations encompass a more restricted range of both,
$\Delta V_{\rm ecl}$ and $V$ than those of Groot et al.\ (1998).
Considering the magnitude of the comparison star to GS~Pav (UCAC4 102-102986)
of $V = 14.24$ (taken from the NOMAD catalogue; Zacharias et al. 2005) 
or $V = 14.51$ (SPM4 catalogue; Girard et al. 2011),
$\Delta m$ then translates into a $V$ magnitude of GS~Pav which ranges 
between 
$14^{\raisebox{.3ex}{\scriptsize m}}_{\raisebox{.6ex}{\hspace{.17em}.}}8$ and
$15^{\raisebox{.3ex}{\scriptsize m}}_{\raisebox{.6ex}{\hspace{.17em}.}}5$,
while $\Delta V_{\rm ecl}$  is restricted to 
$1^{\raisebox{.3ex}{\scriptsize m}}_{\raisebox{.6ex}{\hspace{.17em}.}}8$ --
$2^{\raisebox{.3ex}{\scriptsize m}}_{\raisebox{.6ex}{\hspace{.17em}.}}3$ (see 
Table~\ref{eclipse depth}). This corresponds to what Groot et al.\ (1998) call 
the ``shallow branch'', where the eclipse depth decreases with decreasing 
out of eclipse brightness. In the context of their entire $\Delta V_{\rm ecl}$
-- $V$ diagram they interpret the shallow branch as a range where an
increase of the accretion disk radius conspires with a partial 
self-occultation of a flared disk to decrease $V$ while $\Delta V_{\rm ecl}$
also decreases. Here, no clear indication of such an effect is seen. 
Possibly, the conspiration between disk radius variations and self-occultation
(disk flaring may be a function of azimuth in a time dependent manner) did
not lead to a clear dependence of $\Delta V_{\rm ecl}$ on $V$ in 
2016\footnote{Note also that the linear $\Delta V_{\rm ecl}$ -- $V$ relationship 
on the shallow branch of Groot et al.\ (1998) depends on only 4 data points 
and thus may be statistically fragile.}.

\subsection{Out-of-eclipses variations}
\label{GS Out-of-eclipse variations}

Flickering in GS~Pav appears to occur on a similar amplitude scale as in
EC~21178-5417. However, since GS~Pav is significantly fainter than 
EC~21178-5417 (which has $V=13.65$ according to 
Zacharias et al. 2013) 
data noise makes it even more difficult to quantitatively investigate 
flickering here and I refrain to do so.

\begin{figure}
   \parbox[]{0.1cm}{\epsfxsize=14cm\epsfbox{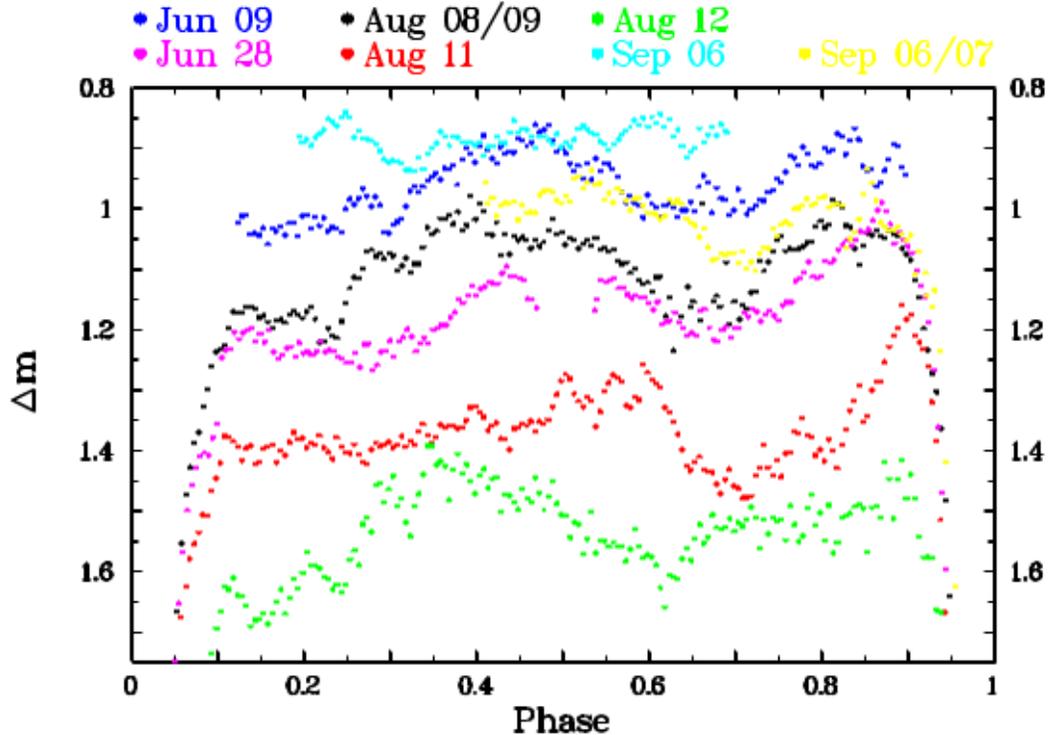}}
      \caption[]{Phase folded out-of-eclipse light curves of GS~Pav
                 (binned in intervals of width 0.005) on the original
                 differential magnitude scale [except for the data of 2016,
                 August 12 (green dots) which have been shifted upwards by
                 $0^{\raisebox{.3ex}{\scriptsize m}}_{\raisebox{.6ex}{\hspace{.17em}.}}3$].
                 (For visualization of the colours used in this figure, the
                 reader is referred to the web version of this article.)}
\label{gspav-ooe}
\end{figure}

Most of the light curves exhibit a more or less consistent pattern of
out-of-eclipse variablity as a function of orbital phase. 
This can be seen in Fig.~\ref{gspav-ooe} 
where all light curves with a suitable phase coverage are shown folded
on the orbital period (on their original differential magnitude scale in
order to preserve the long term variations of the system 
brightness\footnote{Expect for the light curve of 2016, August 12 (green 
dots), which has been shifted upwards by 
$0^{\raisebox{.3ex}{\scriptsize m}}_{\raisebox{.6ex}{\hspace{.17em}.}}3$
for better use of the figure space.}).
In order to better visualize the phase dependent variations and to reduce
the noise the light curves have been binned in intervals of width 0.005.

All light curves which cover the respective phase exhibit an increase in
brightness just before eclipse ingress. This is the usual manifestation
of a hot spot at the location of impact of the stream of transferred matter
onto the accretion disk, as seen in numerous CVs with high orbital
inclination. Apart from this, a~second hump is seen, the phase of which is
not constant. Its maximum occurs roughly between phase 0.3 and 0.6. The 
amplitude is comparable to that of the hot spot hump. This
feature is only missing in the light curve obtained on 2016, September 6
(uppermost curve in Fig.~\ref{gspav-ooe}) when the system was in a particularly
bright state. As in EC~21178-5417 the persistent phase dependent variations
point at azimuthal structures in the accretion disk.

In order to investigate the presence of other than these obvious 
periodicities I combined light curves observed close in time
(set \#1: June 9 and 10; set \#2: August 8/9, 11 and 12; set \#3: 
September 6 and 6/7)
after removal of the eclipses and subtraction of the average nightly 
differential magnitude in order to remove night-to-night variations.
Lomb-Scargle periodograms where then calculated. As expected in view of the
phase folded light curves of Fig.~\ref{gspav-ooe}, they are
dominated by strong signals at twice the orbital frequency and its aliases.
However, a different picture emerges after removal of variations on longer
times scales ($>$0.05 days). This was achieved by applying a Fourier filter
to the light curves which removes variations more rapid than this, and 
subsequent subtraction of the filtered version from the original light curve.

\begin{figure}
   \parbox[]{0.1cm}{\epsfxsize=14cm\epsfbox{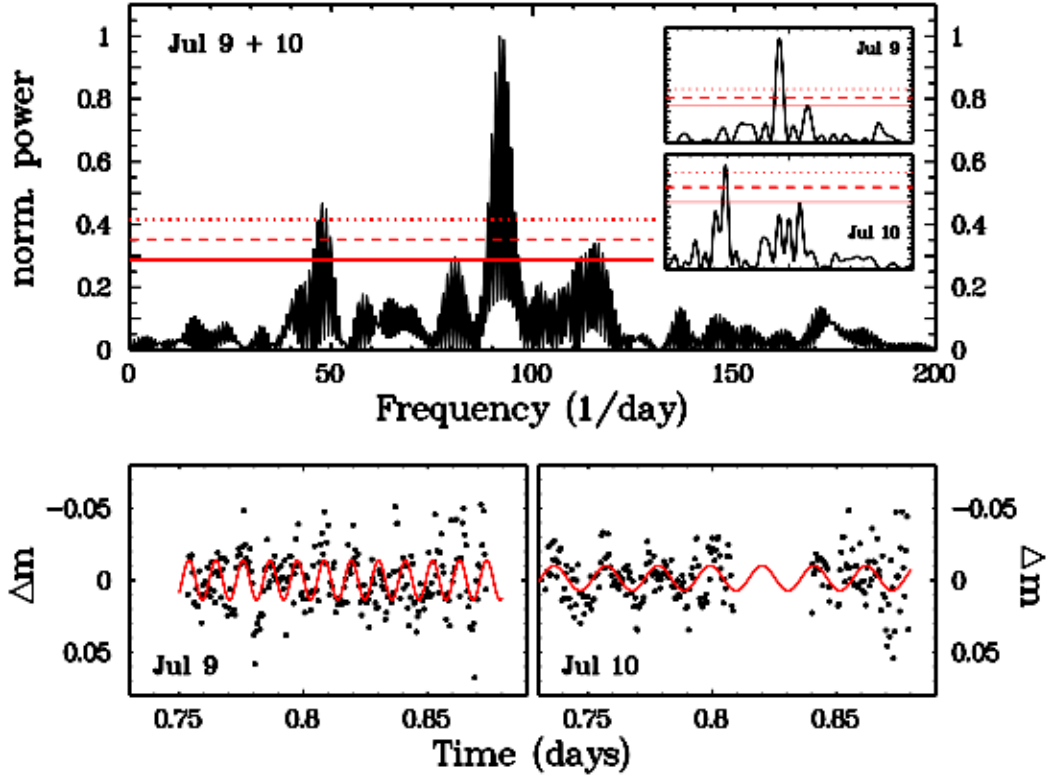}}
      \caption[]{{\it Top:} Power spectrum of the combined light curves of
                 GS~Pav on 2016, June 9 and 10 after removing variations
                 on time scales above 0.05 days and binning in intervals of
                 $\approx$43 sec (0.0005 days). The inserts contain the
                 powerspectra of the light curves of the individual nights.
                 The red solid, dashed and dotted horizontal lines indicate 
                 the power level for false alarm probabilities of 0.01, 0.001
                 and 0.0001, respectively.
                 {\it Bottom:} The light curves of GS~Pav on 2016, June 9 
                 (left) and 10 (right) used to calculate the power spectra
                 in the upper panel. The red curves are best fit sine waves
                 with periods fixed to values corresponding to the highest
                 peaks in the respective power spectra.
                 (For visualization of the colours used in this figure, the
                 reader is referred to the web version of this article.)}
\label{gspav-osc}
\end{figure}

Power spectra of the resulting data show some interesting features which are
most clear-cut in the case of data set \#1. The resulting Lomb-Scargle 
periodogram is shown in the upper frame of Fig.~\ref{gspav-osc}. It is
dominated by a signal corresponding to a period $P^* = 15.7 \pm 0.5$~min.
A~second, weaker signal occurs at $P^{**} = 30.2 \pm 1.1$~min\footnote{Here, 
the errors were calculated from the width $\sigma$ of a Gaussian fitted to the 
corresponding peaks of the power spectra of light curves restricted to the 
individual nights.}, i.e, at exactly $2P^*$ within the errors.
The significance of the periods can be assessed by the red horizontal lines
in the figure which represent the power level for false alarm probabilities
of (from bottom to top) 0.01, 0.001 and 0.0001, calculated using the recipe
of Bruch (2016). However, the oscillations giving rise to these
signals do not occur simultaneously. This becomes obvious when power spectra
of the light curves of the two contributing nights are calculated separately
(inserts in Fig.~\ref{gspav-osc}): $P^*$ only appears on June 9, $P^{**}$ only
on June 10. The light curves are plotted in the lower frame of 
Fig.~\ref{gspav-osc}, where the red curves represent best fit sine curves
with periods fixed to $P^*$ and $P^{**}$, respectively. The periods are thus
not stable from night to night, but the exact commensurability of 
$P^*$ and $P^{**}$ is intriguing. Transient variations of similar nature
appear to be present also in the light curves of other systems. A beautiful
example is the novalike cataclysmic variable 
HQ~Mon (see Fig.~10 of Bruch \& Diaz 2017). 

During other nights GS~Pav exhibits similar behaviour with quasi-periodic
signals occurring in the range of roughly 10 -- 35~min. However, the respective
power spectrum signals are not as significant and the oscillations are not
as coherent as those observed on June 9 and 10. Therefore, if real at all,
they blend in with flickering activity, constituting yet another example of
the difficulty to distinguish between an accidental superposition of random 
brightness fluctuations and short lived quasi periodic variations, as pointed 
out previously by Bruch (2014, 2016).

\subsection{OPOs and DNOs}
\label{GS Pav QPOs and DNOs}

GS~Pav exhibits clear evidence for the occasional presence of QPOs. Although
applying the criterion adopted in the case of EC~21178-5417 that such 
oscillations should be obvious to the eye (see~sect.~\ref{EC DNOs and QPOs}) 
does not reveal them, the technique of stacked power spectra strongly 
suggests the presence in various light curves of oscillations with quasi 
periods of the order of 200 sec -- 500 sec which can persist 
for several hours with some modulation in frequency and amplitude.

\begin{figure}
   \parbox[]{0.1cm}{\epsfxsize=14cm\epsfbox{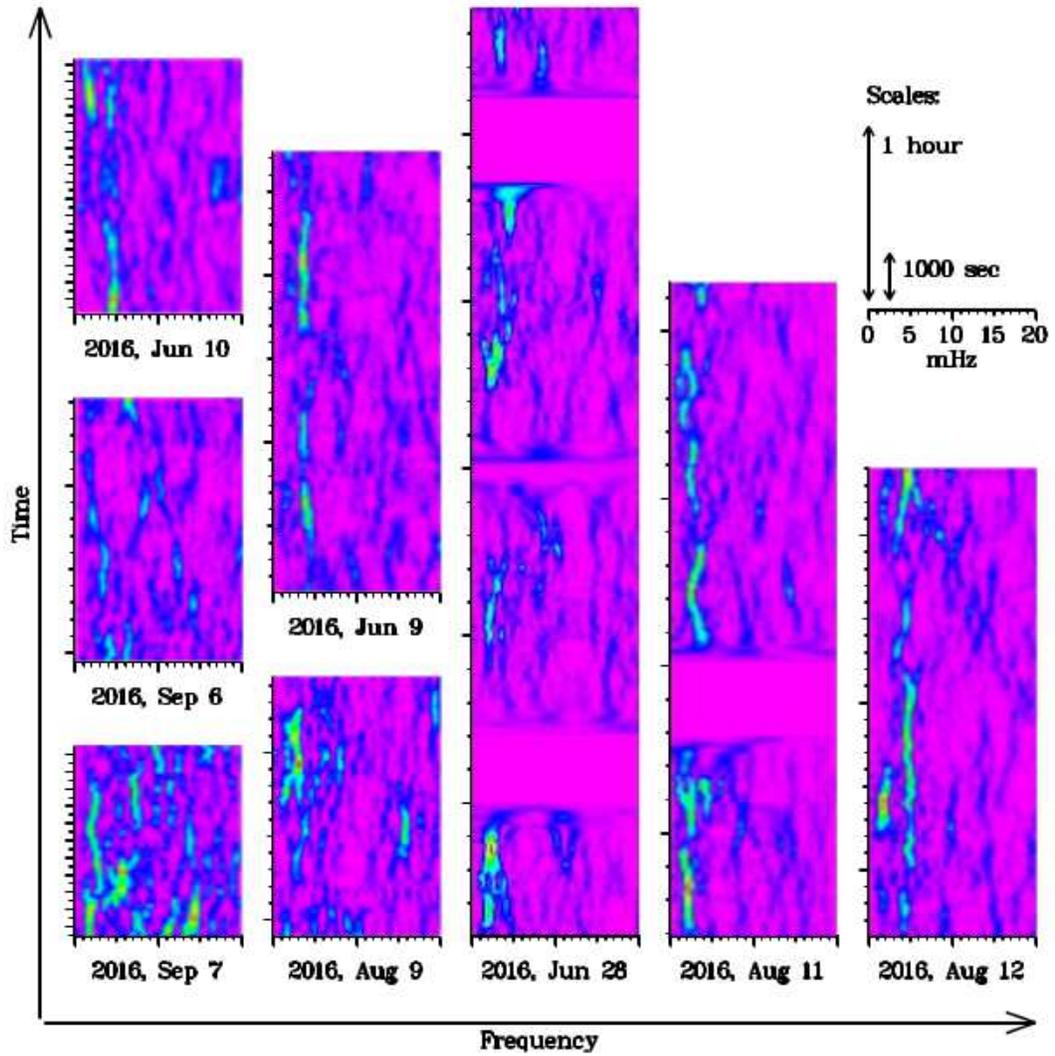}}
      \caption[]{Stacked power spectra GS~Pav in several nights after
                 subtraction of variations on time scales longer than
                 12~min. The spectra are normalized
                 to their nightly maxima and colour coded on a linear scale
                 such that violet and red represent zero and maximum power,
                 respectively. Spectral features within a range
                 of 1000 sec (indicated by the length of the respective
                 double arrow in the upper right of the figure) are not 
                 independent from each other. The horizontal violet bars in
                 the spectra of 2016, June 28 and August 11 represent intervals
                 without data (eclipses or interruptions of the observations).
                 (For visualization of the colours used in this figure, the
                 reader is referred to the web version of this article.)}
\label{gspav-stacked}
\end{figure}

To show this, stacked power spectra of the suitable parts of all observed
light curves of GS~Pav were calculated after removal of variations on time
scales longer than 12~min by subtraction of a Fourier filtered version of
the light curves which only contain modulations on longer time scales. As
in the case of EC~21178-5417 Lomb-Scargle periodograms of
sections of the light curves were constructed, adopting an independence limit
of 1000 sec and allowing for an overlap of
900 sec between subsequent~sections. The resulting stacked power 
spectra for the frequency interval 0 -- 20 mHz are shown in 
Fig.~\ref{gspav-stacked} as a function of time and frequency. Power is
colour coded on a linear scale
such that red represents the maximum values measured during a
night\footnote{A very strong but localized signal at $\approx$2~mHz close
to the end of the light curve of 2016, June 10 (see feature in the upper left
corner of the stacked power spectra of this night) has artificially been
reduced in power by a factor of 2 in order to enhance the visual appearance
of the persistent signal close to 4.5 mHz.}
and violet represent zero power (the horizontal violet bars on 2016, June 26
and August 11 are intervals without data either due to eclipses or interruptions
of the observations). 

During many of the nights conspicuous signals in the form of more or less
continuous vertical streaks in the stacked power spectra in the range of
approximately 2 - 5 mHz (500 - 200~sec), sometimes undulating
somewhat in frequency, extend over periods of time well in excess of the
independence limit. This behaviour
may be most convincing on 2016, August 12 (rightmost spectrum of 
Fig.~\ref{gspav-stacked}) where a signal centred at 4.8 mHz (208~sec)
exends over more than half the light curve (lower part of the 
stacked spectrum) and is possibly resumed later on. On June 9 a signal 
appears with interruptions at 3.8 mHz (263~sec). Before the eclipse
on August 11 there appears to be an oscillation at 2.3 mHz (435~sec),
and immediately after the eclipse an oscillations which undulates between
3.9 mHz (256~sec) and 2.5 mHz (400~sec) can be discerned. Whether 
the signals later
in that light curve should be considered as a continuation of the undulating
oscillation or are due to isolated events is not clear. On June 10, the light
curve starts with a significant signal at 4.7 mHz (212~sec), 
then appears to split into two branches, each of them being fainter
than the original signal, only to join into a single oscillations of
4.4 mHz (227~sec), similar to the original one, at the end of the 
observations. During other nights, similar signals in the same frequency
range appear, but is not as clear whether they represent QPOs or are just
due to isolated flickering events.

Warner (2004) pointed out that QPOs are often observed in combination\
with DNOs. However an investigation of the high frequency part of the stacked
power spectra of GS~Pav (beyond the upper frequency limit shown in 
Fig.~\ref{gspav-stacked}) did not reveal convincing evidence for DNOs.
This is not surprising since the S/N ratio of the respective light
curves is not better that of the EC~21178-5417 data where in spite of the
detection by Warner et al.\ (2003) of a plethora of DNOs none could be seen
(unless drawing on pre-information; not available in the case of GS~Pav)
possibly due to the insufficient S/N ratio. 

\section{V345~Pav}
\label{V345 Pav}

Most of what is known about V345~Pav (= EC~19314-5915) comes from a study of
Buckley et al.\ (1992). They presented a detailed spectroscopic analysis,
supplemented by some photometry. Based on eclipses observed in the latter
they derived an orbital period of 4.7543~h. The spectrum of V345~Pav
is peculiar in the sense that it exhibits metallic absorption lines typical 
of a G8 dwarf star which cannot be attributed to the secondary of the CV
system. Instead, Buckley et al.\ (1992) conjectured the presence of a third
star. As in the case of GS~Pav, the initial motivation to include V345~Pav in
the present study was the expectation to significantly improve the precision
of the orbital period through observations of additional eclipses. 

\subsection{Eclipse timings and ephemeris}
\label{V345 Eclipse timings and ephemeris}

The minimum times of the 14 eclipses observed in V345~Pav were determined
in the same way as for EC~21178-5417 and GS~Pav. 
The typical error turned out to be 16.2~sec. The eclipse times 
are listed in Tab.~\ref{Eclipse timings} where 
eclipse no.\ \#0 is arbitrarily assigned to the eclipse observed on 2015,
May 22. These data permit to calculate preliminary ephemeris with
sufficient precision to bridge the gap in time to the observations of
Buckley et al.\ (1992) without cycle count ambiguity (negative eclipse
numbers). As was the case
for GS~Pav, Buckley et al.\ (1992) list their eclipse timings of V345~Pav
in HJD. {\it Assuming} that they also used the UTC time scale, I transformed
them into BJD-TDB and include them in Table~\ref{Eclipse timings}.
Combining all eclipse timings then enables to calculate
long-term ephemeris. The eclipses epochs of V345~Pav are given by:
\begin{eqnarray}
{\rm BJD_{min}} & = & 2457164.79433\ (20) \nonumber \\ [-0.5ex]
               &    & + 0.1980963877\ (28) \times E \nonumber 
\end{eqnarray}
The rms-error of the residuals between
the observed eclipse times and those calculated from these ephemeris
is 23.8~sec. Again, the $O-C$ values for the individual eclipses
are included in Table~\ref{Eclipse timings}.

\subsection{Eclipse profile}
\label{V345 Eclipse profile}

\begin{figure}
   \parbox[]{0.1cm}{\epsfxsize=14cm\epsfbox{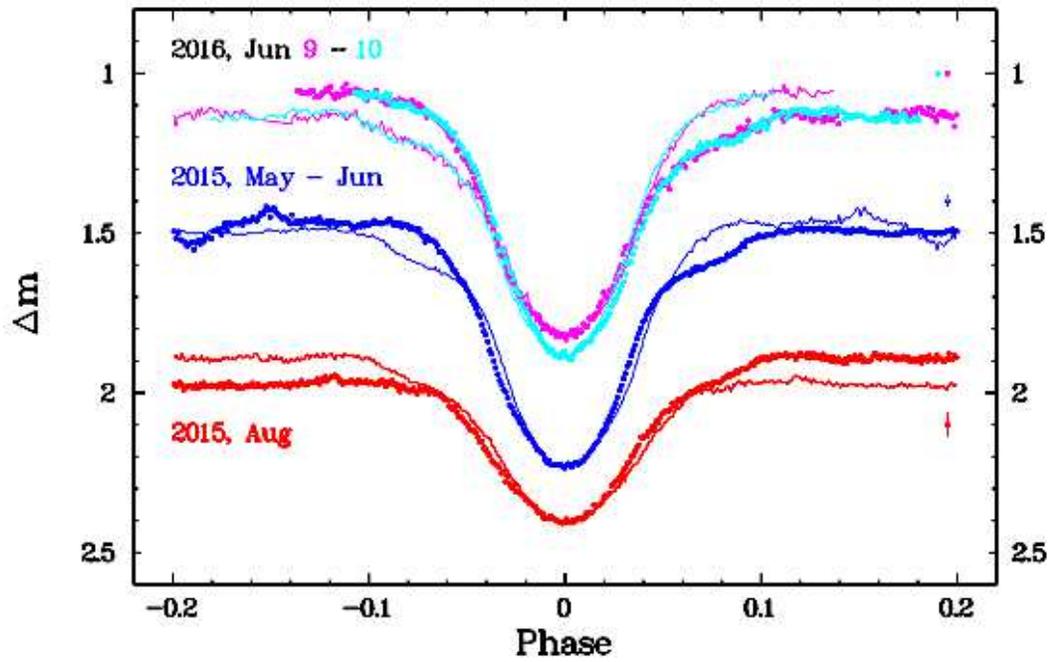}}
      \caption[]{Eclipse profiles of V345~Pav: eclipses of 2016, 
                 June 9 (magenta) and 10 (cyan), both shifted upward by 
                 $0^{\raisebox{.3ex}{\scriptsize m}}_{\raisebox{.6ex}{\hspace{.17em}.}}3$
                 for clarity; average of five eclipses of
                 2015, May - June (blue); average of seven eclipses
                 of 2015, August (red), shifted downward by 
                 $0^{\raisebox{.3ex}{\scriptsize m}}_{\raisebox{.6ex}{\hspace{.17em}.}}3$.
                 For details of the organization of the figure,
                 see Fig.~\ref{ec-eclprof}. 
                 (For visualization of the colours used in this figure, the
                 reader is referred to the web version of this article.)}
\label{v345-eclprof}
\end{figure}

The light curves of V345~Pav can be separated into three distinct groups. The
first one includes the observations of 2015 May and June, the second those
of 2015, August, and the final one those of 2016, June. 
Fig.~\ref{v345-eclprof} (which is organized in a similar way
as Fig.~\ref{ec-eclprof}) shows the eclipse profiles of the three epochs. 
The graphs corresponding to 2015, August and 2016, June have been shifted 
downwards and upwards, respectively, by 
$0^{\raisebox{.3ex}{\scriptsize m}}_{\raisebox{.6ex}{\hspace{.17em}.}}3$ for clarity. 
While the eclipses of the first two epochs have been averaged (five eclipses
of 2015, May -- June and seven eclipses of 2015, August), the 2016, June
eclipses are shown individually in order to visualize the significant 
difference in eclipse depth from one night to the next, while the 
out-of-eclipse level of the two light curves remained remarkably stable.

The deviations from symmetry of the eclipse profile is more pronounced in
V345~Pav (in particular in 2015, May -- June) than was observed in
EC~21178-5417 and GS~Pav but are qualitatively similar, again exhibiting
a delayed hot spot egress. Additionally, the 2015 eclipses contain
asymmetries in their mid-sections, when egress is steeper than ingress. 
Only the lower part of the eclipses appear perfectly symmetric.

The eclipse amplitudes, as gauged from the difference between the 
average magnitude in the phase intervals $-0.16 < \phi < -0.14$ and 
$0.14 < \phi < 0.16$\footnote{Since these phases were not covered in the 2016, 
June light curves slightly different intervals had to be used in 
these nights.} and the minimum of a high order polynomial fit to the
eclipse bottom, are listed in Table~\ref{eclipse depth}.
The significantly larger scatter of the 2015, August eclipse depths 
(eclipse numbers 407 -- 428) indicates 
that the system was in a much less stable state than during May-June when
the eclipse depth was very stable. 

%

The eclipses are shallower than in the other two targets of this study. 
This is due to the presence of the 
third star in the system. Buckley et al.\ (1992) estimate that it contributes
34\% of the out-of-eclipse light in the $V$ band. Since
the isophotal wavelength of our white light observations roughly 
correspond to $V$ (Bruch 2017) the expected eclipse depth in the
absence of the third component can be calculated. 
Assuming that the percentage contribution of its light is constant (i.e.,
the out-of-eclipse variations are all due to uncertainties in the zero point
of the differential magnitude scale between missions; see 
Sect.~\ref{Observations and data reductions}) 
the average depth 
during the three epochs of observation would then be 
$1^{\raisebox{.3ex}{\scriptsize m}}_{\raisebox{.6ex}{\hspace{.17em}.}}53$,
$0^{\raisebox{.3ex}{\scriptsize m}}_{\raisebox{.6ex}{\hspace{.17em}.}}82$ and
$1^{\raisebox{.3ex}{\scriptsize m}}_{\raisebox{.6ex}{\hspace{.17em}.}}55$.

\subsection{Out-of-eclipses variations}
\label{V345 Out-of-eclipse variations}

Flickering in V345~Pav reaches only about half the amplitude that it attains
in the other two targets of this study and is thus quite weak compared to most
CVs. Again, data noise inhibits a more
details quantitative investigation. In order to study the phase dependence of
the out-of-eclipse variations, as in the previous cases the eclipses
were first removed from the light curves. I distinguish
between the 2015, May -- June data and the light curves obtained in 2015, 
August. The data of June 12 were excluded from this analysis because they are
much noisier than the rest, and those of August 10 because only a small part
of the out-of-eclipse phases was observed. In order to remove slight 
night-to-night variations the average magnitude was subtracted from the
individual nightly light curves which were then combined into two data
sets, comprising the nights of May -- June and of August, respectively.
Finally, these data sets were folded on the orbital period.

\begin{figure}
   \parbox[]{0.1cm}{\epsfxsize=14cm\epsfbox{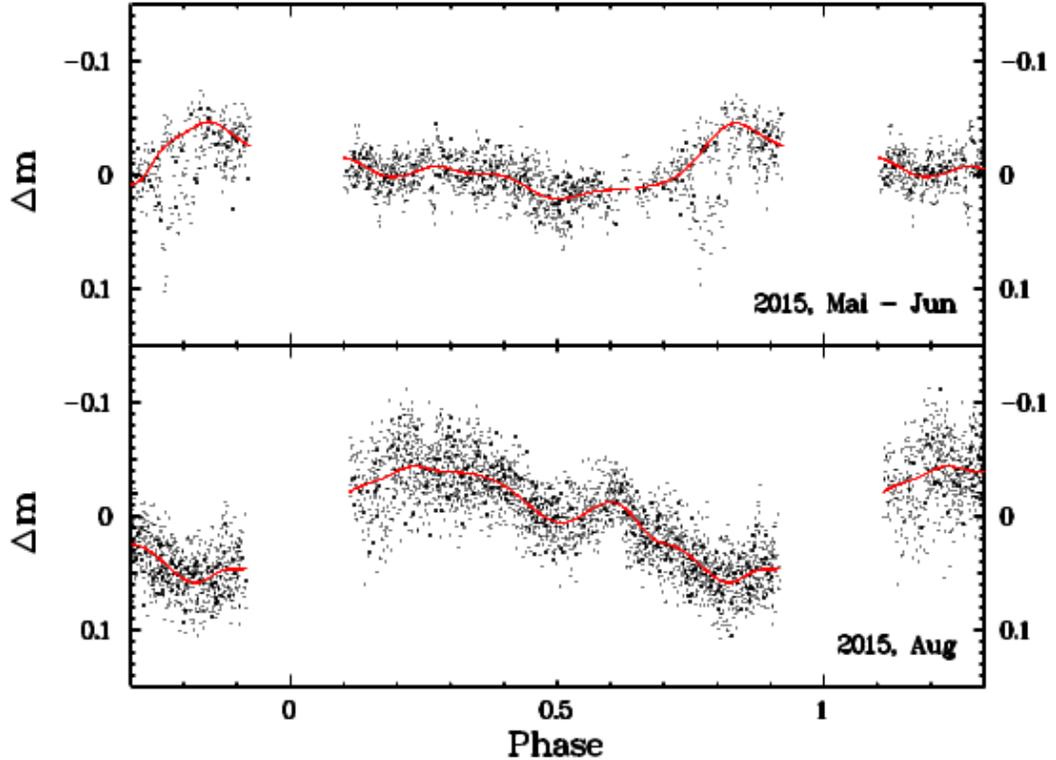}}
      \caption[]{Phase folded out-of-eclipse light curves of 
                 V345 Pav of 2015, May -- June (top) and August 
                 (bottom). The red lines are smoothed versions of the
                 original data. 
                 (For visualization of the colours used in this figure, the
                 reader is referred to the web version of this article.)}
\label{v345-out-of-eclipse}
\end{figure}

The results are shown in 
Fig.~\ref{v345-out-of-eclipse}. The 2015, May -- June light curves appears
to contain a faint orbital hump just before the eclipse at a
phase when the classical hot spot is expected to rotate into view in 
cataclysmic variables\footnote{The points deviating to fainter magnitudes
between phase 0.75 -- 0.85 are all part of the light curve of 2015, May 22
which has a minimum at those phases but then recovers to the average
hump maximum magnitude.}. After the eclipse, the magnitude declines
gradually to a minimum just before the onset of the hump.
In contrast, in August V345~Pav attains an asymmetrically
shaped maximum (the rise is steeper than the decline) roughly at phase 
$\phi = 0.25$ and a minimum close to $\phi = 0.82$, with the decline
being interrupted by a smaller secondary
maximum close to $\phi = 0.6$. The increase in brightness
across the eclipse is also clearly seen in Fig.~\ref{v345-eclprof}. Note
also that in agreement with the clear presence of the orbital hump in
May-June and its absence in August the delayed hot spot eclipse egress is 
more pronounced in the first than in the second of these epochs (see
Fig.~\ref{v345-eclprof}). The overall time dependent behaviour of the 
out-of-eclipse variations again suggests the presence of an accretion disk with
variable azimuthal structure leading to aspect dependent modulations in this
high inclination system. The changing phase of the light curve maximum may 
serve as a warning against the perils to derive an orbital period (in the
absence of eclipses) from repetitive (but only apparently periodic) 
brightness modulations alone.

In order to verify the presence of periodic variations other than the orbital
modulation Lomb-Scargle periodograms of the combined May-June and August
light curves were calculated. They only contain sinals at the orbital
frequency, its first overtone and their one/day aliases.

\subsection{DNOs and QPOs}
\label{V345 DNOs and QPOs}

In contrast to EC~21178-5417, no previous information about the presence of
DNOs, lpDNOs or QPOs is available for V345~Pav. I searched for such 
oscillations in the out-of-eclipse light curves in the same way as has been
done in Sects.~\ref{EC DNOs and QPOs} and \ref{GS Pav QPOs and DNOs}. 
While, again, the stacked power spectra
contain a myrad of signals, none of them attains a false alarm probability
low enough, extends beyond the independence limit, or repeats itself in \
different nights to claim it as real with any degree of confidence. 
Therefore, I conclude that if oscillations are present in the data they 
remain below the detection limit. This is in agreement with
Buckley et al.\ (1992) who also did not find evidence for DNOs or QSOs.

\section{Conclusions}
\label{Conclusions}

Time resolved photometry of three eclipsing novalike variables was
investigated in order to (i) determine or to refine orbital ephemeris,
(ii) quantify the properties of the eclipses, (iii) characterize the
out-of-eclipse variations, and (iv) search of QPOs and DNOs. The longer
time base over which the objects were observed, and -- in the case of 
GS~Pav and V345~Pav -- combining them with published observations taken
decades earlier, permitted to increase the precision of the period by
one (GS~Pav) to three (EC~21178-5417 and V345~Pav) orders of magnitude.

In all cases the eclipse profile is clearly structured, exhibiting a 
delay in late eclipse egress, i.e., the familiar feature caused by the
egress of the hot spot on the accretion disk. Otherwise, the eclipses
are symmetrical to a high degree. There depth exhibits significant variations
on the time scale of weeks (i.e., between observing missions), but at least
in GS~Pav, variations occur also on shorter time scales. The same is true
for the light level just before or after eclipse. In GS~Pav, the correlation
between out-of-eclipse brightness and eclipse depth observed by
Groot et al.\ (1998) could not be confirmed. In view of the difficulty to
measure precise contact phases (in particular during eclipse ingress) I
resist the temptation to determine dynamical and geometric system properties
using eclipse profiles.

Out of eclipse the variations remain modest but show some systematics. In all
target stars flickering occurs only on a low scale with amplitudes restricted
to $\sim$$0^{\raisebox{.3ex}{\scriptsize m}}_{\raisebox{.6ex}{\hspace{.17em}.}}1$
in EC~21178-5417 and GS~Pav and 
$\sim$$0^{\raisebox{.3ex}{\scriptsize m}}_{\raisebox{.6ex}{\hspace{.17em}.}}05$
in V345~Pav. Superposed are systematic variations on
longer time scales with a single maximum between phase 0.5 and 0.7 and an
amplitude of 
$\approx$$0^{\raisebox{.3ex}{\scriptsize m}}_{\raisebox{.6ex}{\hspace{.17em}.}}25$
in EC~21178-4517, or a hump between
phase 0.3 and 0.6 in addition to the classical hot spot hump just before
eclipse (amplitudes: 
$0^{\raisebox{.3ex}{\scriptsize m}}_{\raisebox{.6ex}{\hspace{.17em}.}}1$ --
$0^{\raisebox{.3ex}{\scriptsize m}}_{\raisebox{.6ex}{\hspace{.17em}.}}2$) in GS~Pav. In 
V345~Pav the available data suggest that the pattern of out-of-eclipse 
variations is less stable on time scales of weeks or months. 
Apart from the described orbital variations at least in two nights GS~Pav
exhibits a periodicity on shorter time scales during time intervals of 
several hours. In one night a period of 15.7~min was measured,
while in the next night it doubled (within the measurement error) to
30.2~min. 

A multitude of QPOs and DNOs has been observed by Warner et al.\ (2003) in
EC~21178-5417. These could not be retrieved in the present observations, 
probably because of a lower S/N ratio of the data. Only during one night
(drawing on pre-information about the properties of these oscillations)
possible indications for their presence were identified. A search for
similar signals in V345~Pav was to no avail. However, in some light curves
of GS~Pav clear signs of persistent signals with somewhat modulated frequencies
(corresponding to periods between 200 and 500~sec) and amplitudes were
detected in stacked power spectra.  
 
\section*{Acknowledgements}

I gratefully acknowledge the use of observations from the AAVSO International
Database contributed by observers worldwide.

\section*{References}

\begin{description}
\parskip-0.5ex

\item
             Beckemper, S. 1995, Statistische Untersuchungen zur St\"arke des
             Flickering in kataklysmischen Ver\"anderlichen,
             Diploma thesis, M\"unster
\item
             MIRA: A Reference Guide (Astron.\ Inst.\ Univ.\ M\"unster
\item
             Bruch, A. 2014, A\&A, 566, A101
\item
             Bruch, A. 2016, New Astr., 46, 60
\item
             Bruch, A. 2017, New Astr., in press 
\item
             Bruch, A., Diaz, M.P. 2017, New Astr., 50, 109
\item
             Buckley, D.A.H., O'Donoghue, D., Kilkenny, D., Stobie, S.R., \&
             Remillard, R.A. 1992, MNRAS, 258,285
\item
             Deeming, T.J. 1975, Ap\&SS, 39, 447
\item
             Eastman, J., Siverd, R., \& Gaudi, B.S. 2010, PASP, 122, 935
\item
             Girard, T.M., Van Altena W.F., Zacharias, N., et al. 
             2011, AJ, 142, 15
\item
             Groot, P.J., Augusteijn, T., Barziv, O., \& van Paradijs, J. 1998,
             A\&A 340, L31
\item
             Hoffmeister, C. 1963, Ver\"off.\ Sternw.\ Sonneberg, 6, 1
\item
             Lomb, N.R. 1976, ApSS, 39, 447
\item
             Ritter, H., \& Kolb, U. 2003, A\&A, 404, 301
\item
             Scargle, J.D. 1982, ApJ, 263, 853
\item
             Schwarzenberg-Czerny, A. 1989, MNRAS, 241, 153
\item
             Stellingwerf, R.F. 1978, ApJ, 224, 953
\item
             Stobie, R.S., Kilkenny, D., O'Donoghue, D., et al. 
             1997, MNRAS, 287, 848
\item
             Warner, B. 2004, PASP, 116, 115
\item
             Warner, B., Woudt, P.A., \& Pretorius, M.L. 2003, MNRAS, 344, 1193
\item
             Zacharias, N., Monet, D.G., Levine, S.E., et al.
             2005, AAS, 205, 4815
\item
             Zacharias, N., Finch, C.T., Girard, T.M., et al. 2013, AJ, 145, 44
\item
             Zwitter, T., \& Munari, U. 1995, A\&AS, 114, 575

\end{description}

\end{document}